\documentclass[pra,showpacs,twocolumn,floatfix]{revtex4-1}

\usepackage{graphicx}
\usepackage{verbatim}
\usepackage{comment}
\usepackage{amsmath}
\usepackage{amssymb}
\usepackage{stmaryrd}
\usepackage{color}

\begin{document}

\title{Thermodynamics of strongly interacting fermions in two-dimensional optical lattices}

\author{Ehsan Khatami and Marcos Rigol} 
\affiliation{Department of Physics, Georgetown University, Washington DC, 20057 USA}
\affiliation{Kavli Institute for Theoretical Physics, University of California, Santa Barbara, 
Santa Barbara, California 93106, USA}

\begin{abstract}
We study finite-temperature properties of strongly correlated fermions in 
two-dimensional optical lattices by means of numerical linked cluster expansions,
a computational technique that allows one to obtain exact results in the 
thermodynamic limit. We focus our analysis on the strongly interacting regime, 
where the on-site repulsion is of the order of or greater than the band width. 
We compute the equation of state, double occupancy, entropy, uniform susceptibility, 
and spin correlations for temperatures that are similar to or below the ones 
achieved in current optical lattice experiments. We provide a quantitative analysis 
of adiabatic cooling of trapped fermions in two dimensions, by means of both 
flattening the trapping potential and increasing the interaction strength.
\end{abstract}

\maketitle

\section{Introduction}
\label{sec:intro}

Recent optical lattice experiments have opened a new venue for exploring the effects of 
strong correlations in quantum lattice models. For example, the superfluid to Mott-insulator 
transition for bosons has been observed in geometries of three \cite{m_greiner_02},
two \cite{i_spielman_07}, and one \cite{t_stoferle_04} dimension. 
Currently, there is a race to access temperatures low enough for the transition to the 
antiferromagnetically ordered Ne\'{e}l state in three dimensions, or possibly more exotic 
states in two dimensions, to be observed for fermions~\cite{r_jordens_08,u_schneider_08}. 
So far, the interaction strength and the temperature in lattice fermion experiments remain
relatively high in comparison to the hopping amplitude $t$. This is in part because $t$, 
which is set by optical lattice parameters, is in general small in the regimes where 
one-band models are applicable.

On the theoretical side, there is an ever-increasing demand for precise numerical results 
for the relevant parameters of the Hubbard model and for large enough system sizes, which could 
be used to interpret current experiments and also provide suggestions for future 
experiments~\cite{r_hekmes_08a,t_paiva_10,r_jordens_10,e_gorelik_10,s_fuchs_11,s_chiesa_11}.
For this model, especially for strong interactions, the present computations become 
particularly challenging as the temperature is lowered below the hopping amplitude. 

Here, we study various thermodynamic quantities such as the equation of state, 
entropy, double occupancy, and spin correlations in the thermodynamic limit for
interactions up to three times the band width, utilizing numerical linked cluster 
expansions (NLCEs)~\cite{M_rigol_06,M_rigol_07b}. We obtain a detailed 
understanding of the evolution of various quantities with adiabatically increasing  
interaction strength, of great interest to current optical lattice experiments. 
Using the local density approximation (LDA), we analyze the thermodynamics of fermions 
in a harmonic trap and calculate their temperature as a function of the interaction 
strength and total entropy. We also present a quantitative analysis of various cooling 
schemes for the experiments \cite{l_deleo_08,tl_ho_09,f_heidrich_09}.

\section{model}

We consider the two-dimensional (2D) Hubbard Hamiltonian,
\begin{equation}
\hat{H}=-t\sum_{\left <i,j\right >\sigma}(\hat{c}^{\dagger}_{i\sigma} \hat{c}_{j\sigma} + \text{H.c.})+
U\sum_i \hat{n}_{i\uparrow} \hat{n}_{i\downarrow}+\sum_{i\sigma} V_i \hat{n}_{i\sigma},
\label{eq:ham}
\end{equation}
where $\hat{c}^{\dagger}_{i\sigma}$ ($\hat{c}_{i\sigma}$) 
creates (annihilates) a fermion with spin $\sigma$ on site $i$, and 
$\hat{n}_{i\sigma}=\hat{c}^{\dagger}_{i\sigma} \hat{c}_{i\sigma}$ is the number operator. 
$\left<..\right>$ denotes nearest neighbors (NNs), $U$ is the strength of the on-site 
repulsive interaction, and $V_i$ is a space-dependent local chemical potential. 
$t=1$ ($\hbar=1$ and $k_B=1$) sets the energy scale throughout this paper.

\section{computational approach}

In linked-cluster expansions \cite{linked}, we express an extensive property of the 
model per lattice site in the thermodynamic limit ($P$) in terms of contributions from 
all the clusters, up to a certain size, that can be embedded in the infinite lattice:
\begin{equation}
P=\sum_c L(c)w_p(c),
\label{eq:1}
\end{equation}
where $c$ represents the clusters. This contribution is proportional to the weight
of each cluster for that property [$w_p(c)$] and to its multiplicity [$L(c)$]. The latter
is defined as the number of ways in which that particular cluster can be embedded in the 
infinite lattice, per site. The weight, on the other hand, is calculated recursively as the 
property for each cluster [$\mathcal{P}(c)$] minus the weights of all its subclusters:
\begin{equation}
\label{eq:2}
w_p(c)=\mathcal{P}(c)-\sum_{s\subset c}w_p(s).
\end{equation}
Here, we use the NLCE, where $\mathcal{P}(c)$ is computed by means of full exact 
diagonalization~\cite{M_rigol_06}. Because of the exact treatment of individual clusters 
in the NLCE, the series converge at significantly lower temperatures in comparison to 
high-temperature expansions in which perturbation theory is used~\cite{M_rigol_06}.

NLCEs are complementary to quantum Monte Carlo (QMC) approaches, such as the determinantal 
QMC (DQMC)~\cite{r_blankenbecler_81}, or dynamical mean-field theory~\cite{a_georges_96} 
and its cluster extensions, such as the dynamical cluster approximation (DCA)~\cite{m_hettler_98,m_jarrell_01}. 
They can also help to benchmark future experiments as well as new computational techniques. 
This is because NLCEs do not suffer from statistical or systematic errors, such as 
finite-size effects, and, as opposed to the DQMC and DCA, they are not restricted to small or 
intermediate interaction strengths. In Ref.~\cite{supp.}, we make our raw NLCE data 
for a wide range of interactions available for comparison. 

The validity of NLCEs, however, is limited to a region in temperature in which the series
converge (the convergence region). We have found that, for the Hubbard model, NLCEs converge 
down to lower temperatures as the strength of the interaction is increased. At half-filling, 
and for interactions larger than the band width, NLCEs can access the region with strong 
antiferromagnetic (AF) correlations, identified by the suppression of the 
uniform susceptibility. Although, the method does not have any systematic restriction away 
from half-filling, in the latter region, the series fail to converge at temperatures as low 
as those accessible to the half-filled case. This prevents us from accessing low-temperature
phases, such as $d$-wave superconductivity, that arguably exist in this model at finite doping.

We begin our analysis with the homogeneous system ($V_i=0$) in the grand canonical ensemble. 
For each $U$, we compute all properties for a very dense grid of chemical potential ($\mu$)
and temperature, so that we can also follow properties at constant density ($n$) \cite{M_rigol_07b}. 
The NLCE calculations are carried out on the square lattice up to the ninth order in the site 
expansion (nine sites). We use Wynn and Euler algorithms for summing the terms in the series 
to extend the region of convergence~\cite{M_rigol_06}. Since only NN hopping is considered, 
all properties of the particle-doped system can be expressed in terms of those
from the hole-doped system. Hence, away from half-filling, we only show results for 
the hole-doped system.

\section{results}

\subsection{Equation of State}

\begin{figure}[!t]
\centerline {\includegraphics*[width=3.3in]{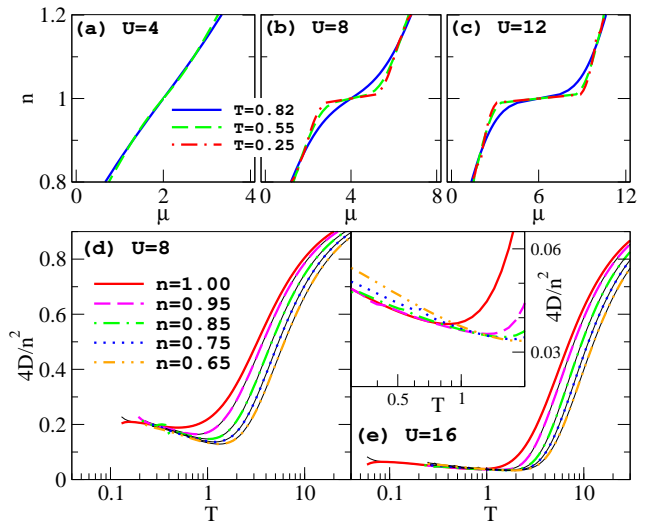}} 
\caption{(Color online) Top: Equation of state for (a) $U=4$, (b) $U=8$, 
and (c) $U=12$ and at three different temperatures. Except for $U=4$ at $T=0.25$, 
NLCE results converge for all the values of chemical potential presented here. 
Only the last order of the series is shown after using Wynn sums with three cycles 
of improvement. Bottom: Normalized double occupancy vs $T$ at four hole dopings 
for (d) $U=8$ and (e) $U=16$. We use Euler sums for the last six terms at 
half-filling and Wynn sums for $n\ne1$. Thin (black) lines in (d) and (e) are the 
results for the one to last order of NLCEs after the above sums. The inset in (e) 
magnifies the low-temperature region for $U=16$. The unit of energy is set to the 
hopping amplitude $t$.}
\label{fig:state}
\end{figure}

The equation of state for the Hubbard model provides important information about 
correlation effects as the strength of the on-site interaction is increased, and 
can be studied in optical lattice experiments. In Figs.~\ref{fig:state}(a)-(c), 
we depict the equation of state at three different temperatures, $T=0.82$, $0.55$, 
and $0.25$, for the weak-, intermediate-, and strong-coupling regimes ($U=4$, $8$, 
and $12$, respectively). For the last two values of $U$ [Figs.~\ref{fig:state}(b) 
and \ref{fig:state}(c)], one can see the emergence of an incompressible region 
around $\mu=U/2$, a clear signature of the Mott gap opening in the density of states 
at low temperatures. 

\subsection{Double Occupancy}


In Figs.~\ref{fig:state}(d) and \ref{fig:state}(e), we show the double occupancy, 
$D=\langle \hat{n}_{\uparrow} \hat{n}_{\downarrow} \rangle$, normalized by its uncorrelated 
high-temperature value ($n^2/4$) for $U=8$ and $16$, respectively. The double occupancy 
exhibits a clear low-$T$ rise with decreasing temperature. This feature has attracted a lot 
of attention recently, especially after the real-space DMFT study of the three-dimensional 
(3D) version of the model in a harmonic trap~\cite{e_gorelik_10}. Gorelik {\em et al.} argued 
that the onset of the AF ordering in the strong-coupling regime is signaled by an enhanced double 
occupancy, which can be directly measured in optical lattice experiments. However, according
to Figs.~\ref{fig:state}(d) and \ref{fig:state}(e), the low-temperature rise occurs not only 
at half-filling, but also away 
from it. Moreover, the rise starts at even higher temperatures for higher dopings. This 
implies that the enhancement of $D$ in the trap upon lowering the temperature is significant 
in the Mott-insulating core as well as in other areas of the trap where density is $<1$. 
Therefore, in real experiments, such an enhancement can be observed in systems that 
have a very small or even no Mott insulating region at the center of the trap. Hence, the
observation of an increase in $D$ alone may not signal the onset of AF order. To ensure that
AF order is emerging, one must also make sure that the density is $1$ in most of the trap.

For large values of $U$ [see, e.g., $U=16$ in Fig.~\ref{fig:state}(e)], the normalized $D$ is 
almost independent of doping below $T\sim 1$ and down to the lowest accessible temperatures
for $n\gtrsim 0.85$ (see inset), implying that $D\propto n^2$ in this region. One can understand 
the latter from the fact that local moments are likely ordered, and the double occupancy arises 
from virtual hoppings to NN sites, so a relatively small number of extra holes only modifies 
the probability of those hoppings (accounted for by $n^2$), not the actual process.

\subsection{Entropy}


Generally, when using QMC-based methods, entropy calculations involve numerical
derivatives and/or integration by parts~\cite{a_dare_07,k_mikelsons_09}, which can introduce 
systematic errors. Within NLCEs, the entropy is computed directly from its definition in the grand 
canonical ensemble: 
\begin{equation}
 S=\ln(Z)+\frac{\langle \hat{H}\rangle -\mu \langle \hat{n}\rangle }{T},
\end{equation}
where $Z$ is the partition function.

\begin{figure}[!t]
\centerline {\includegraphics*[width=3.35in]{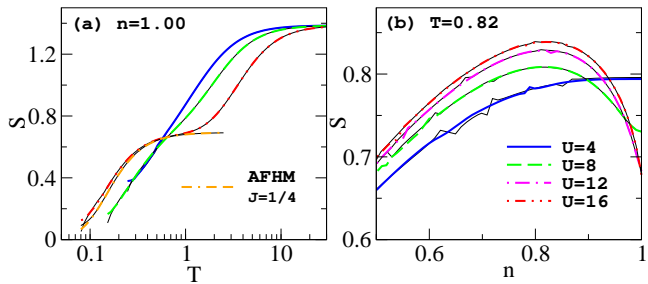}} 
\caption{(Color online) Entropy (a) vs $T$ at half-filling, and (b) vs $n$ at $T=0.82$ for 
different values of $U$. Thick (thin) lines are results for the last (one to last) order 
of the expansions as explained in Fig.~\ref{fig:state}. In (a), we have also included the entropy 
for the AFHM with the exchange interaction $J=0.25$.}
\label{fig:EN}
\end{figure}

We first study the entropy at half-filling. Results are shown in Fig.~\ref{fig:EN}(a) 
as a function of the temperature for $U=4$, $8$, and $16$. There are two distinct 
regions of fast decrease in the entropy in the strong-coupling regime, e.g., $U=16$. 
Those regions are separated by a crossing point of curves for different values of $U$ 
around $T=0.6$, corresponding roughly to $S=\ln(2)$. The emergence of these two regions 
results from the fact that as $U$ increases, charge degrees of freedom are suppressed 
at higher temperatures due to the higher price of double occupancy, 
and at the same time, the characteristic energy scale of the spin degrees of freedom, 
$J=4t^2/U$, becomes smaller, pushing the low-$T$ drop to lower temperatures. 
We find that in the latter region, the entropy curves for large $U$ ($\gtrsim 14$) 
follow very closely the entropy of the antiferromagnetic Heisenberg model (AFHM). 
This is shown for $U=16$ in Fig.~\ref{fig:EN}(a), where we also plot the entropy of 
the AFHM with $J=0.25$. 

In Fig.~\ref{fig:EN}(b), we show the entropy away from half-filling for a range of 
interactions at a fixed $T=0.82$. In the weak-coupling regime (e.g., $U=4$), the 
entropy increases monotonically with the density and is maximal at half-filling. Since 
correlations play a small role, the system behaves similarly to a noninteracting 
system; i.e., the closer to the point where there is an equal number of electrons 
and holes, the higher the entropy. This trend changes upon increasing $U$, for which 
the moment ordering suppresses the entropy significantly close to half-filling. 
As a result, there is a maximum in the entropy in the vicinity of $n\sim 0.85$ 
for all interactions in the strong-coupling regime~\cite{J_bonca_03}. Below, we 
discuss how these features are reflected in the properties of trapped systems.

\begin{figure}[!t]
\centerline {\includegraphics*[width=3.35in]{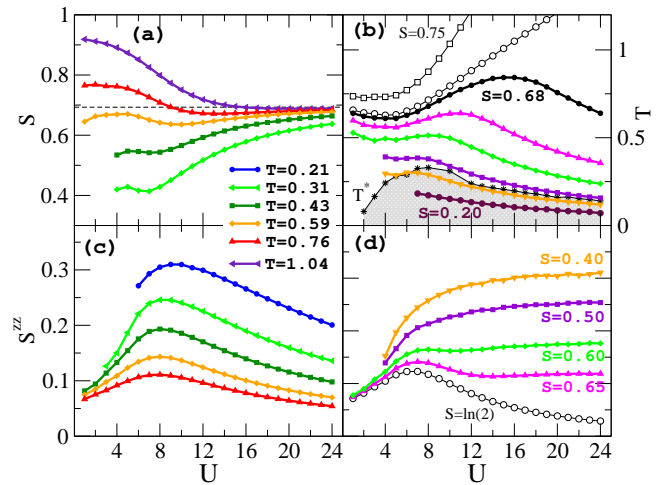}} 
\caption{(Color online) Interaction dependence of different quantities at half-filling. 
Left: (a) Entropy and (c) NN spin correlations at fixed temperatures. Right: 
(b) Temperature and (d) NN spin correlations at constant entropies. In (b), $T^*$ 
represents a crossover temperature to the region where AF correlations grow exponentially 
with decreasing temperature (shaded area).}
\label{fig:SP}
\end{figure}

\begin{figure}[!t]
\centerline {\includegraphics*[width=3.35in]{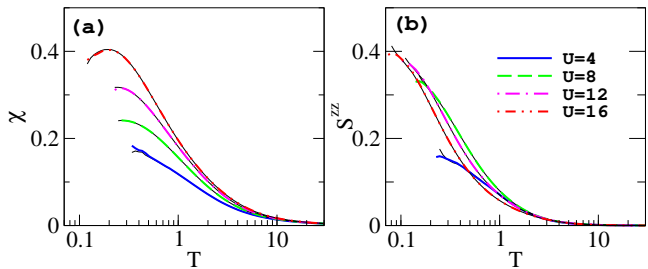}} 
\caption{(Color online) (a) Uniform spin susceptibility and (b) NN spin correlations 
at half-filling vs temperature for different interactions. 
$\chi$ peaks at $T^*$, below which AF correlations grow exponentially with decreasing 
temperature. $S^{zz}$ also shows a sharp increase around $T^*$.}
\label{fig:SS}
\end{figure}

We further take advantage of the fact that, within NLCEs, arbitrary values of $U$ can be 
studied at no additional computational cost and determine the dependence of the quantities 
of interest on $U$. In Fig.~\ref{fig:SP}(a), we show $S$ at half-filling as a function of 
$U$ at fixed temperatures. The temperature regions identified for the entropy in 
Fig.~\ref{fig:EN}(a) are more clearly seen in Fig.~\ref{fig:SP}(a); i.e., the entropy stays 
more or less the same for $T\sim0.6$ (the crossing point) as $U$ increases, whereas it 
generally decreases (increases) with $U$ for $T>0.6$ ($T<0.6$).

The possibility of adiabatic cooling with increasing $U$ in optical lattices has been studied 
by a number of groups recently~\cite{f_werner_05,a_dare_07,t_paiva_10}. Considering interactions
that are often no larger than $3/2$ of the band width, they argue that this process 
is rather weak in two dimensions. Here, we revisit the problem by including results for values 
of $U$ up to three times the band width [see Fig.~\ref{fig:SP}(b)]. We find that if we start
at relatively high values of entropy below $\ln(2)$ (e.g., $S=0.68$), accessible to current 
2D experiments, and continue to increase $U$, the half-filled system can be cooled down to 
very low temperatures. For lower entropy, e.g., $S=0.4$, one can even access the region with 
exponentially large AF correlations below the crossover temperature, $T^*$, as shown in 
Fig.~\ref{fig:SP}(b). We take $T^*$ as the temperature where the uniform spin susceptibility 
($\chi$) as a function of temperature peaks~\cite{t_paiva_10}. We find that $T^*$ also coincides 
with the temperature at which the NN spin correlations, 
$S^{zz}=|\langle \sum_{\langle ij\rangle} S^z_iS^z_j\rangle|$, show rapid growth with decreasing 
temperature. The uniform spin susceptibility and the NN spin correlations are depicted in 
Fig.~\ref{fig:SS}. For $U\gg 1$, $T^*$ is expected to scale with the AF exchange constant 
($J$) in the effective Heisenberg model, i.e., $\propto 1/U$. This behavior, which is demonstrated
in Fig.~\ref{fig:SP} (b), is observed not only for $T^*$, but also for the large-$U$ tails of 
the isentropic curves when $S<\ln(2)$, as $J$ is the only energy scale in that region. For $U<12$ 
in the weak-coupling regime, $T^*$ is not accessible to our NLCE. Therefore, we have taken $T^*$ 
from the DQMC results in Ref.~\onlinecite{t_paiva_10} for that region. It is interesting to 
see that this crossover temperature peaks when the value of the interaction is around the 
band width.

\begin{figure}[t]
\centerline {\includegraphics*[width=3.35in]{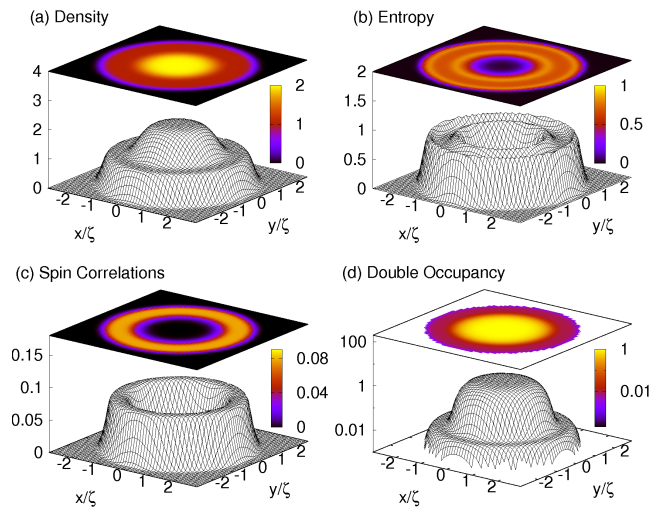}}
\caption{(Color online) (a) Density,  (b) entropy, (c) NN spin correlations, and (d) 
double-occupancy profiles of fermions in a harmonic trap with $\tilde{\rho}=22.9$, 
governed by the Hubbard model with $U=16$ at $T=0.76$. The average entropy per particle 
is $s=0.56$. $\zeta=(2dt/V)^{1/2}$ is the characteristic length.}
\label{fig:3D}
\end{figure}

\subsection{Nearest-neighbor Spin Correlations}
\label{sec:NN}


What is perhaps more important from the experimental point of view 
is how AF correlations change during the process of adiabatically increasing $U$.
As mentioned in Sec.~\ref{sec:intro}, one of the current main goals in cold fermion 
experiments is to achieve AF in the Mott-insulating state. However, the challenge in 
this case lies not only in realizing such a state but also in detecting it. 
Very recently, experimental breakthroughs have been reported which allow the detection 
of NN spin correlations~\cite{s_Trotzky_10,d_greif_11}. 

NN spin correlations, $S^{zz}$, can also be computed exactly using NLCEs. As expected, 
we find that $S^{zz}$ is largest at half-filling for all interactions. Therefore, we 
focus on the half-filled system and plot this quantity per site vs $U$ at constant 
temperatures in Fig.~\ref{fig:SP}(c) and at constant entropies in Fig.~\ref{fig:SP}(d).
The dependence of $S^{zz}$ on $U$ at constant $T$ is nontrivial. As the temperature 
is lowered to $T\sim0.3$, a peak develops in the spin correlations around $U=8$, 
which is indicative of the largest effective exchange interaction between NN spins. 
The peak is a result of the interplay between weak moment formation in the weak-coupling 
regime ($U<8$) and the $1/U$ decrease in the effective $J$ in the  strong-coupling regime. 
We find that at lower temperatures ($T=0.21$), the maximum of $S^{zz}$ occurs at $U\sim9$, 
which is not expected to change significantly with further decreasing temperature.

\begin{figure}[!t]
\centerline {\includegraphics*[width=3.3in]{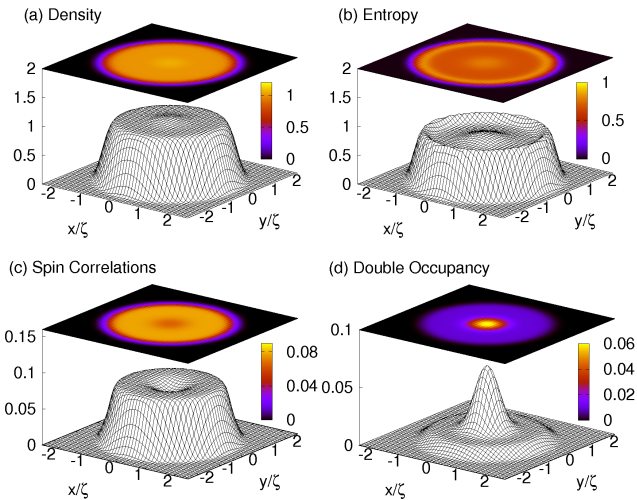}}
\caption{(Color online) Same as Fig.~\ref{fig:3D}, but for $\tilde{\rho}=10.8$. The average
entropy per particle is $s=0.85$ in this case.}
\label{fig:3D2}
\end{figure}

At constant entropy, on the other hand, this picture is strongly modified. Figure \ref{fig:SP}(d) 
shows that $S^{zz}$ saturates to a finite entropy-dependent value with increasing 
$U$ along the isentropic paths in Fig.~\ref{fig:SP}(b), provided that $S<\ln(2)$. 
Note that, even though adiabatic cooling may not be efficient to arrive at regions with large AF 
correlations in 2D~\cite{a_dare_07}, the value of the NN spin correlations will be maximal 
in the large-$U (\gg 12)$ region if $S<0.6$. This is convenient for 
experiments in optical lattices for which $U$ is typically large compared to the band width.

\subsection{Trapped Systems}


\begin{figure}[b]
\centerline {\includegraphics*[width=3.35in]{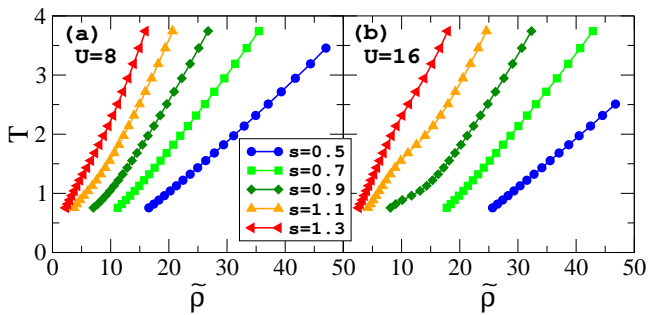}}
\caption{(Color online) Temperature vs characteristic density at constant entropies per 
particle for (a) $U=8$ and (b) $U=16$.}
\label{fig:Isen}
\end{figure}

To make direct contact with experiments in optical lattices, we study the manifestation 
of our previous results in systems confined by a spatially varying harmonic potential, 
$V_i=Vr_i^2$. Here, $r_i$ denotes the radial distance of each site to the center of the 
trap, and for any given value of $U$, all properties of the system are determined by 
the characteristic density $\tilde{\rho}=N (V/2dt)^{d/2}$~\cite{m_rigol_03}, where $d$ 
is the dimensionality and $N$ is the number of particles. The resulting inhomogeneous 
Hubbard model is then studied using the LDA along with our results for the infinite 
system. A recent QMC study of the inhomogeneous Hubbard model \cite{s_chiesa_11} 
has shown that the LDA is a good approximation for local observables at the temperatures 
accessible here. We should stress that NLCEs are ideal for this kind of study
because, for each value of $U$, one can compute all properties for a very dense grid of 
temperatures and chemical potentials at almost no additional computational cost. The same 
is, of course, not true for QMC-based calculations, where each temperature and chemical 
potential requires a separate computation.

In Fig.~\ref{fig:3D}(a), we plot the resulting density profile for $U=16$ at $T=0.76$. 
We have chosen $\tilde{\rho}=22.9$ such that there are band-insulating ($n=2$) and 
Mott-insulating ($n=1$) domains in the trap. Very useful information for the experiments
is provided by the spacial distribution of the density, entropy, NN spin correlations, 
and double occupancy, as shown in Fig.~\ref{fig:3D}. The entropy 
is minimal ($0$) in the band insulator, peaks at $n\sim1.18$ and $0.82$, consistent with Fig.~\ref{fig:EN}(b), 
and has a local minimum in the Mott ring. In the latter region, spin correlations are maximal, and
as expected for this large value of $U$, the double occupancy is large only in the region where $n>1$.

In Fig.~\ref{fig:3D2}, we show the same quantities as in Fig.~\ref{fig:3D}, for the 
reduced $\tilde{\rho}$ of $10.8$ at the same temperature and interaction strength.
As a result of this isothermic change, the entropy per particle increases from $0.56$ 
to $0.85$. Nevertheless, the Mott-insulating region with a relatively uniform entropy
profile is clearly seen over most of the trap.

It has become apparent in experiments with fermions in optical lattices that cooling 
approaches beyond the standard evaporative cooling techniques are required if one is to 
reach temperatures low enough that exotic physics emerges. Three recent proposals 
have shown how to generate low-entropy states where a large fraction of the system 
is in a band-insulating domain (i.e., with large values of $\tilde{\rho}$) 
\cite{j_bernier_09,tl_ho_09,f_heidrich_09}. The idea is then that one can adiabatically 
reduce the trap strength (the characteristic density $\tilde{\rho}$) so that the 
effective temperature of the fermions decreases. In this way, antiferromagnetism and other 
low-temperature phenomena can be explored.

\begin{figure}[t]
\centerline {\includegraphics*[width=3.35in]{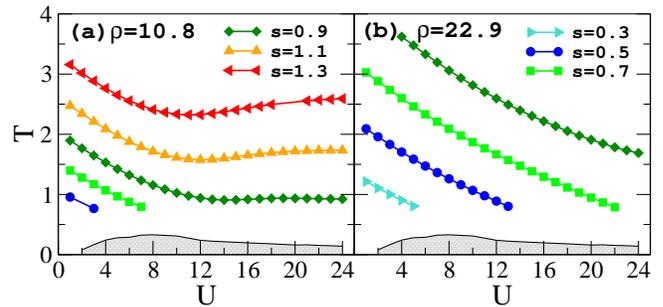}}
\caption{(Color online) Temperature vs $U$ at constant entropies per 
particle for (a) $\tilde{\rho}=10.8$ and (b) $\tilde{\rho}=22.9$. The shaded area, the same as in 
Fig.~\ref{fig:SP}(b), is the region of exponentially large AF correlations below 
$T^*$ in the Mott insulating core of the trap.}
\label{fig:Isen_TU}
\end{figure}

In Fig.~\ref{fig:Isen}, we show quantitatively 
how this idea works for trapped 2D systems. We plot the temperature as a function
of $\tilde{\rho}$ for various values of the total entropy per particle. Recent studies have 
shown that the entropy per particle ($s$) for a particular $U$ can be estimated by fitting 
the double-occupancy measurements at different $\tilde{\rho}$ to data from numerical 
simulations~\cite{r_jordens_10}. In Figs.~\ref{fig:Isen}(a) and \ref{fig:Isen}(b) one can 
see that, for the two values of $U$ shown, the temperature decreases rapidly with 
decreasing $\tilde{\rho}$, demonstrating that this approach works very efficiently 
for 2D trapped systems. The inflection point, shown, e.g., for $s=0.9$
in Fig.~\ref{fig:Isen}(b), is the signature of a large Mott region forming in the 
trap (as seen in Fig.~\ref{fig:3D2}). This occurs provided the entropy is low enough and for a 
range of characteristic densities that depends on $U$. AF ordering in the Mott core
emerges at $T^*$, which, for low entropies, can be reached before the Mott insulator 
is destroyed by further flattening of the trap. 

It is also interesting to study what happens to the temperature of a trapped system 
as one increases the interaction strength at constant entropy. [Results for homogeneous systems 
at half-filling are presented in Fig.~\ref{fig:SP}(b).]  In Fig.~\ref{fig:Isen_TU}, 
we show isentropic curves in the $T-U$ plane for trapped systems at various entropies 
and for the two characteristic densities, $\tilde{\rho}=10.8$ and $22.9$, 
used in Figs.~\ref{fig:3D} and \ref{fig:3D2}. As expected from the results in Fig.~\ref{fig:Isen}, 
the shape and location of the isentropic curves depend strongly on the value of $\tilde{\rho}$.
In Fig.~\ref{fig:Isen_TU}, we also show the same shaded area as in Fig.~\ref{fig:SP}(b)
below $T^*$, which, here, represents the region where the Mott-insulating core of the trap 
develops large AF correlations. Our calculations show that cooling can take place 
in trapped systems as the interaction increases. The entropies at which cooling is
observed, and the values of $U$ at which cooling occurs, depends on the characteristic 
density in the trap. Hence, as reported in Ref.~\cite{t_paiva_11} for 3D systems, 
adiabatically increasing the interaction strength can allow experimentalists to reach 
the temperatures needed to observe the onset of (quasi-)long-range AF 
correlations in a trapped system. Unfortunately, unlike for the homogeneous system at
half-filling, our NLCEs do not provide access to the temperatures relevant to that region 
for the 2D trapped system. 

\section{Summary}

In summary, utilizing NLCEs, which, within the convergence temperature region are 
free of statistical and/or systematic errors and provide exact results in the thermodynamic 
limit, we have calculated thermodynamic properties, such as the equation of state, double 
occupancy, entropy, uniform susceptibility, and NN spin correlations, 
of the 2D Hubbard model for a wide range of interaction strengths and temperatures. 
Precise data for the entropy on a dense temperature grid allowed us to study temperature 
and NN spin correlations, relevant to optical lattice experiments, as a function of 
the entropy. We find that for any $S<\ln(2)$, by adiabatically increasing $U$ 
to very large values, the temperature decreases as $1/U$ and the spin correlations 
saturate to an entropy-dependent value beyond $U\sim 12$. Using the LDA, we have discussed 
the implications of our results for lattice fermions in the presence of a confining 
harmonic potential. In particular, we have shown how cooling can be achieved by reducing 
the confinement strength in a system that starts with a wide band-insulating 
domain in the center of the trap, or by adiabatically increasing the interaction 
strength.

\section*{Acknowledgments}

This work was supported by NSF under Grants No.\ OCI-0904597 and No.~PHY05-51164.
We thank A. Muramatsu, R. T. Scalettar, R. R. P. Singh, and K. Mikelsons for useful 
discussions.

\end{document}